\documentclass[12pt]{article}
\def\+{{+\!\!\!+}}
\def\pp{\mbox{\tiny${}_{\stackrel\+ =}$}}
\def\mm{\mbox{\tiny${}_{\stackrel= \+}$}} 
\def\cA{{\cal A}}

\def\cM{{\cal M}}
\def\cN{{\cal N}}
\def\fb{{\textstyle{1\over{(3-2N)}}}}
\def\d{\partial}
\def\bd{{\bar\partial}}
\def\a{\alpha}
\def\th{\theta}
\def\g{\gamma}
\def\G{\Gamma}
\def\bG{\bar\Gamma}
\def\N{\nabla}
\def\bN{\bar\nabla}
\def\tb{{\bar\theta}}

\def\de{\delta}
\def\P{\Psi}
\def\f{\varphi}
\def\bfi{{\bar\varphi}}
\def\bP{{\bar\Phi}}
\def\h{\chi}
\def\bh{{\bar\chi}}
\def\r{\rho}
\def\br{{\bar\rho}}
\def\l{\lambda}
\def\bl{{\bar\lambda}}
\def\L{\Lambda}

\def\s{\sigma}
\def\p{\psi}
\def\bp{{\bar\psi}}
\def\e{\epsilon}
\def\eb{{\bar \epsilon}}

\def\implies{\Rightarrow}
\def\is{\equiv}

\def\bF{{\bar F}}

\def\um{{\underline{m}}}
\def\un{{\underline{n}}}
\def\up{{\underline{p}}}
\def\ua{{\underline{a}}}
\def\ub{{\underline{b}}}
\def\uc{{\underline{c}}}

\def\hN{\hat{\nabla}}   
\def\hbN{\hat{\bar{\nabla}}}    

\def\half{{\textstyle{1 \over 2}}}
\def\ihalf{{\textstyle{i \over 2}}}
\def\fou{{\textstyle{1 \over 4}}}

\def\bop#1{\setbox0=\hbox{$#1M$}\mkern1.5mu
\vbox{\hrule height0pt depth.04\ht0
\hbox{\vrule width.04\ht0 height.9\ht0 \kern.9\ht0 \vrule
width.04\ht0}\hrule height.04\ht0}\mkern1.5mu}

\newcommand\hv[1]{(\ref{#1})}

\begin{document}

\newcommand{\inv}[1]{{#1}^{-1}} 

\newcommand{\beq}{\begin{equation}}
\newcommand{\eeq}[1]{\label{#1}\end{equation}}
\newcommand{\ber}{\begin{eqnarray}}
\newcommand{\eer}[1]{\label{#1}\end{eqnarray}}
\begin{center}
\hfill USITP-98-12\\
\hfill OSLO-TP 7-98\\
\hfill August 1998\\

\vskip .3in \noindent

\vskip .1in

{\large \bf {2D supergravity in p+1 dimensions }} \vskip .2in

{\bf Henrik Gustafsson}$^{a,}$\footnote{e-mail: henrik@physto.se}
 and {\bf Ulf Lindstr\"om}$^{a,b,}$\footnote{e-mail: ul@physto.se}
\bigskip \\
$\mbox{}^{a)}$ {\em Institute of Theoretical Physics, University of
Stockholm \\
Box 6730,
S-113 85 Stockholm SWEDEN}\\
\bigskip

$\mbox{}^{b)}$ {\em Institute of Physics, University of Oslo\\ Box 1048,

N-0316 Blindern, Oslo, NORWAY\\} \vskip .15in

\vskip .1in
\end{center}
\vskip .4in
\begin{center} {\bf ABSTRACT } \end{center} \begin{quotation}\noindent
We describe new $N$-extended $2D$ supergravities on a $(p+1)$-dimensional
(bosonic) space.
The fundamental objects are moving frame densities that
equip each $(p+1)$-dimensional point with a $2D$ ``tangent space''.
The theory is
presented in a $[p+1, 2]$ superspace.
For the special case of
$p=1$ we recover the $2D$ supergravities in an unusual form. The
formalism
has been developed with applications to the string-parton picture of
$D$-branes at strong coupling in mind. \end{quotation}
\vfill
\thispagestyle{empty}\eject
\setcounter{page}{1}

\section{Introduction}

Diffeomorphism invariant models where the ``gravity'' fields
$e_{\ua}^{\ \um}$ may be
non-invertible moving frames arise in several different contexts.
One example is the Chern-Simons description of $3D$ gravity discussed by
Witten \cite{witt1} and others \cite{3dgr}. In this paper we are
concerned
with the case when $e_{\ua}^{\ \um}$ relate the $(p+1)$-dimensional
space-time
manifold coordinatized by $\xi^{\um}$ to a lower $d$-dimensional
``tangent space''. Such a situation arises in the description of the
tensionless, $T\to 0$ limit of the fundamental bosonic \cite{kl},
supersymmetric \cite{lst2}, and spinning string
\cite{lst1}. It has also been discussed more recently in the context
of a strong coupling
limit of $D$-branes \cite{lind1}.

The $T\to 0$ limit of the spinning string has been given a superspace
description in
terms of a ``null'' superspace \cite{lr} where a $2D$ supergravity based
on non-invertible
$e_{\ua}^{\ \um}$'s is introduced.
This corresponds to $p=1$ and $d=1$ above. For the limit of $D$-branes
the
relevant bosonic dimensions are $p+1$ and $d=2$, and it is this case
which
shall concern us below.

Using the bosonic description in \cite{lind1} as our starting point, we
construct
superspace supergravities based on a superspace with $p+1$ bosonic and
$2N$
fermionic coordinates\footnote{To discriminate between $(p,q)$
superspace,
($p$ left movers and $q$ right movers), and our superspace, we denote the
latter by $[p+1,2]$.}. The basic fields transform as densities and the
space-time field content of the superfields is reduced via constraints.
These constraints take a form which is unfamiliar from the usual $2D$
supergravity point of view, but one which generalizes that used in
\cite{lr}.
Following standard superspace supergravity procedures, we use a
Wess-Zumino
gauge to
display the physical content of the model. In this gauge we solve the
Bianchi
identities and find the component relations that determine the vector
derivative components in terms
of the spinor derivative ones. The component transformations are found
from
the superspace
ones, both for the supergraity fields and for scalar matter fields.
Finally,
as examples,
we present
$\s$-models based on this supergravity for certain
$N$. In fact, the requirements on the Lagrangian limit the number of
supersymmetries to $N=1,2$, if the full superspace measure is used.
Some of these models are expected to be
relevant for supersymmetrization of the models that describe the strong
coupling limit of
$D$-branes.

The plan of the
paper is as follows: Section 2 contains the basic definitions of our
supergravities.
In Section 3 we present the component relations that follow from solving
the
Bianchi identities and in Section 4 we derive the transformations. The
discussion is exhaustive for $N=1$. For higher $N$ we give the lower
components. Section 6
contains the
$\s$-model actions in superspace as well as in components, and Section 7
contains our conclusions. We have collected some useful superspace
relations
used in our derivations in an Appendix where we also explain our
conventions.

\section{Basics}

In this section we define the new $[p+1, 2]$ superspace supergravity.
It should be
compared to the standard $N=1$, $2D$ superspace supergravity as described
in,
e.g.,
\cite{jim} or \cite{rvnz}, and to \cite{howe,jim2} for higher $N$.

The fundamental supergravity objects are \ber
\N_{i\pm} &=&
E_{i\pm}^{\ \ \um}\d_{\um}
+E_{i\pm}^{\ \ j+}\d_{j+}+E_{i\pm}^{\ \ j-}\d_{j-}+\omega_{i\pm} M\cr
&\is&E_{i\pm}^{\ \ \cM}\d_\cM+\omega_{i\pm} M, \eer{nabla}
where $\xi^{\um}$, $\um=0,\ldots,p$
are bosonic coordinates, $\th^{i\pm}$, $i=1,\ldots,N$ are fermionic
coordinates, $\d_{\um}\is\d/\d\xi^{\um}$, $\d_{i\pm}\is\d/\d\th^{i\pm}$,
$M$ is the $2D$ Lorentz generator and
$\cM\in\{\um,i+,i-\}$. Occasionally we will also use the ``tangent
space''
indices
$\cA\in\{\+ ,=,i+,i-\}$. The operators in \hv{nabla} obey the constraints

\ber
&&\{\N_{i+},\N_{j-}\}+\G_{(i+}\N_{j-)}=\delta_{ij}RM,\cr && \ \cr
&&\{\N_{i\pm},\N_{j\pm}\}+\G_{\pm(i}\N_{j)\pm}=\pm 2i\delta_{ij}\N_{\pp},
\eer{scons}
where $R$ is a curvature superfield.
These constraints define the vector derivatives
\beq
\N\pp \is e\pp
^{\um}\d_{\um}+\h\pp^{i+}\d_{i+}+\h\pp^{i-}\d_{i-}+\omega\pp M,
\eeq{vect}
and integration by parts leads to the relations \ber
(1-N)\left(\N_{i\pm}\G_{j\pm}+\N_{j\pm}\G_{i\pm}\right)&=&\pm
i\delta_{ij}
\left(1\cdot {\stackrel \leftarrow \N}\pp\right), \cr \N_{i+}\G_{j-}+
\N_{j-}\G_{i+} &=&0.
\eer{parts}
The ``connection''\footnote{One is inclined to call the corresponding
terms
in \hv{scons} torsion terms, but the density character of the $\N$'s
makes
``connections'' more appropriate.} terms are given by \ber
\G_{i\pm}&\is &\fb\left(\d_{\um} E^{\ \ \um}_{i\pm} +\d_{j+}
E^{\ \ j+}_{i\pm}+\d_{j-}
E^{\ \ j-}_{i\pm}\pm\half\omega_{i\pm}\right)\cr &\is&\fb
\left(1\cdot{\stackrel\leftarrow\N}_{i\pm}\right). \eer{gamma}
All fields are superfields and depend $\xi^{\um},\th^{i+}$ and
$\th^{i-}$. The
$\th$'s transform as (weight
$-\fou$) densities under $\xi$ diffeomorphisms. Diffeomorphisms,
($\s^{\um}$), Supersymmetry, ($\e^{i\pm}$) and Lorentz, ($\L$),
transformations are coded into the superfield $K$ defined by
\beq
K\is \d_{\um}\s ^{\um}+\e^{i+}\d_{i+}+\e^{i-}\d_{i-}+\Lambda M,
\eeq{Kdef}
and the transformations of the derivatives in \hv{nabla} are given by
\ber
\delta \N_{i\pm} &=& [\N_{i\pm} ,K]+\textstyle{1\over{2(N-2)}}
(1\cdot{\stackrel\leftarrow K})\N_{i\pm}, \eer{natra1}
\ber
\delta \N_{\pp} &=& [\N_{\pp} ,K]+\textstyle{1\over{(N-2)}}
(1\cdot{\stackrel \leftarrow K})\N_{\pp}
\eer{natra2}
where
\beq
(1\cdot{\stackrel
\leftarrow K})\is\d_{\um}\s^{\um}-\d_{i+}\e^{i+}-\d_{i-}\e^{i-}.
\eeq{Kback}
These lead to the appropriate transformations for densities of weights
$\fou$
and $\half$
respectively. For $N=2$ we have to constrain the transformations
to be supervolume preserving, i.e. 
$(1\cdot{\stackrel\leftarrow K})=0$.
The relations
\hv{scons},\hv{natra1} and \hv{natra2}
were found by allowing arbitrary coefficients for the density terms, and
matching the resulting component expressions to the bosonic case. Finally
we
mention that the $\G_\pm$'s could be absorbed into fully covariant
derivatives that feel the density character of the objects they act on.
We do that for $N=2$ in Section 5.

\section{Components}

In this section we find the full component content of the $N=1$ theory
and
the first few components for higher $N$.

To display the physical content of the theory it is convenient to work in
a
Wess-Zumino (WZ) gauge which we define as follows: \ber
&& \N_{i\a}|=\d_{i\a}, \cr
&& \left[\N_{i\a}
,\N_{j\beta}\right]|+\G_{i\a}\N_{j\beta}|-\G_{j\beta}\N_{i\a}|=0, \quad
\a,
\beta\in \{ +,- \},
\eer{wz}
where $|$ denotes ``the $\th$-independent part of''.

We define components by projection and use the same notation for the
supergravity superfields and their lowest components: \ber
e\pp^{\um}&\is& e\pp^{\um}|, \qquad \h\pp^{i\a}\is\h\pp^{i\a}| , \qquad
\omega_{\pp}\is\omega_{\pp}|.\cr
R &\is & R|, \qquad \rho_{i\pm} \is \N_{i\pm} R | .
\eer{sgcom}
From \hv{scons} we obtain the relations
\ber
\pm iNE\pp^\cM&=&E_{i\pm}^\cN\left(\d_\cN E_{i\pm}^\cM\right)+
\fb\left(\d_\cN E_{i\pm}^\cN\right) E_{i\pm}^\cM\cr &&\pm\left(\half +
\fb\right)\omega_{i\pm}E_{i\pm}^\cM,\cr \pm i\omega\pp&=&
E_{i\pm}^\cN\d_\cN \omega_{i\pm}+\fb\left(\d_\cN E_{i\pm}^\cN\right)
\omega_{i\pm} .
\eer{cr1}
Using \hv{cr1} and additional relations that follow from \hv{scons} and
\hv{wz} in conjunction with the Bianchi identities we obtain relations
for
the components (in WZ-gauge). The lowest components of the vector
derivative
are determined in terms of the spinor components:
\ber
\d_{i\pm}E_{j\pm}^{\ \ l\pm}| &=&\pm
i\delta_{ij}\h\pp^{l\pm} -\G_{i\pm}|\delta_j^l,\cr \d_{i\pm}
E_{j\pm}^{\ \ l\mp} | &=&\pm
i\delta_{ij}\h\pp^{l\mp} ,\cr
\d_{i\pm} E_{j\mp}^{\ \ l\pm}| &=& 0 ,\cr \d_{i\mp}E_{j\pm}^{\ \ l\pm}|
&=& -\G_{i\mp}|\delta_j^l,\cr \d_{i\pm} E_{j\pm}^{\ \ \um}| &=& \pm
i\delta_{ij}e\pp^{\um} , \cr
\d_{i\pm} E_{j\mp}^{\ \ \um}| &=& 0,\cr
\d_{i\pm}\omega_{j\pm} | &=& \pm
i\delta_{ij}\omega\pp ,\cr
\d_{i\pm} \omega_{j\mp} | &=&
\half\delta_{ij} R ,
\eer{vsrlns}
where
\beq
\G_{i\pm}| = \pm\ihalf \textstyle{1\over{(2-N)}}\delta_{ij}\h\pp^{j\pm}.
\eeq{gAmA}
The level $\th$ relations for the vector derivative components relate
them to
lower ones\footnote{In spite of their seemingly divergent character for
$N=2$,
these relations are applicable for that case too, provided one sets
$\h_\+^{\ \ +}=\h_=^{\ \ -}=0$, see below.}: \ber
\d_{i\pm} e\pp^{\um}|
&=&\pm
i\delta_{ij}\left(\textstyle{{N-1}\over{N-2}}\right)\h\pp^{j\pm}e\pp^{\um
},
\cr \d_{i\pm} e\mm^{\um}| &=&\pm i\delta_{ij}\left( \h\mm^{j\pm}
e\pp^{\um}-
\textstyle{1\over{(2-N)}}\h\pp^{j\pm} e\mm^{\um}\right)\cr \d_{i\pm}
\h\pp^{j\mp} |
&=&\pm
i\left(\textstyle{{N-1}\over{N-2}}\right)\h\pp^{i\pm}\h\pp^{j\mp},\cr
\d_{i\pm} \h\pp^{j\pm} | &=&\de_i^j\left( \half\d_{\um} e\pp^{\um}
\pm\omega\pp
\pm i
\h\pp^{k\mp}\h\mm^{k\mp} \right) \cr
&&\pm
i\left(\textstyle{{N-1}\over{N-2}}\right)\h\pp^{i\pm}\h\pp^{j\pm},\cr
\d_{i\mp} \h\pp^{j\pm} | &=& -\textstyle{i\over{2N}}\de_i^jR\pm
\textstyle{i\over{(2-N)}}
\h\mm^{i\mp}\h\pp^{j\pm}\mp i\h\pp^{i\mp}\h\mm^{j\pm},\cr \d_{i\mp}
\h\pp^{j\mp} | &=&
\de_i^j\left(\half\d_{\um}e\pp^{\um}
\pm i\h\pp^{k\mp}\h\mm^{k\mp}
\right)\cr
&&\pm\textstyle{i\over{(2-N)}}
\h\mm^{i\mp}\h\pp^{j\mp}\mp i\h\pp^{i\mp}\h\mm^{j\mp},\cr
\d_{i\pm}\omega\pp |
&=& \half \h\pp^{i\mp} R\pm
i\left(\textstyle{{N-1}\over{N-2}}\right) \h\pp^{i\pm}\omega\pp ,\cr
\d_{i\mp}\omega\pp | &=&
\left(\textstyle{{2+N(N-2)}\over{2N(N-2)}}\right) \h\pp^{i\pm}R\pm
\textstyle{i\over{(2-N)}}\h\mm^{i\mp}\omega\pp\cr
&&\mp i\h\pp^{i\mp}\omega\mm\pm\textstyle{i\over N}\rho_{i\pm}.
\eer{vth}
The level $\th^2$ spinor derivative components cannot all be determined
for
$N>1$.
For $N=1$ we find:
\ber
\d_+\d_- E_\pm^{\ \pm}| &=& \fou\h_{=}^{\ \ -}\h_{\+}^{\ \ +} -
\half\h_{\+}^{\ \ -}\h_{=}^{\ \ +}, \cr \d_+\d_- E_\pm^{\ \mp} | &=&
-\ihalf
\d_\um e\pp^\um\mp\half\h\pp^\mp\h\mm^\mp , \cr \d_+\d_- E_\pm^{\ \um}|
&=&
\mp\h\mm^\mp e\mm^\um\pm\half\h\mm^\mp e\pp^\um, \cr
\d_+\d_- \omega_\pm | &=& \textstyle{i\over 4}\h\pp^\pm R
\pm\half\h\mm^\mp
\omega\pp\mp\h\pp^\mp\omega\mm\pm\rho_\pm . \eer{sth2}
Using this result and the equations \hv{vsrlns} gives the level $\th$
relations for $\G_\pm$,
\ber
\d_{\pm}\G_{\pm} |&=& \pm \ihalf\left( \d_{\um}e\pp^{\um}\pm \omega\pp
\mp\h\pp^{\mp}\h\mm^{\mp}\right)\cr \d_{\pm}\G_{\mp} |&=&\mp\fou R
\pm\fou
\h_=^{\ \ -}\h_\+^{\ \ +} \mp\half
\h_\+^{\ \ -}\h_=^{\ \ +} .
\eer{hrlns}

The relations \hv{vth} and \hv{sth2} were determined using the lowest
dimension
Bianchi identities. Applying the Bianchi identity to $[\N_\+ ,\N_=]$ for
$N=1$ confirms these relations and leads to the constraint
\beq
\d_{\um}\left(e_{[\+}^{\ \um} e_{=]}^{\ \un}\right)=0. \eeq{Acons}
and an expression for the $\th^2$ component of $R$, \beq
\fou\d_+\d_-R| =-\half
e_{[\+}^{\um}\d_{\um}\omega^{}_{=]}+\omega_\+\omega_=+\G_{[-}\rho_{+]}.
\eeq{Ricci}
Note that $\rho_\pm$ is an independent field. For higher $N$ the
constraint
\hv{Acons} is still valid but new relations for the higher components of
$R$ are found. In particular, $\rho_{i\pm}$ is not independent for $N>1$.

The $\th^2$ components of the vector derivative can be related to lower
ones
using
\hv{scons},\hv{vsrlns} and \hv{sth2}. For $N=1$ these are all the
components.
For higher $N$, the constraints lead to additional relations between
higher
$\th$ components, which we omit.

We shall also need the first few components of a scalar superfield $X$
(in WZ-gauge),
\ber
X&\is& X|, \qquad \Psi_{i\pm}\is\d_{i\pm} X|,\cr {\cal F}_{i\pm
j\pm}&\is&
\d_{i\pm}\d_{j\pm}X|,\qquad {\cal F}_{i\pm j\mp}\is\d_{i\pm}\d_{j\mp}X|.
\eer{Xcom}
Note that the density character of $\th$ leads to $\Psi$ and ${\cal F}$
being
densities.

\section{Transformations}

In this section we present the transformations of the component fields in

WZ-gauge.
To stay in this gauge the transformation superfield $K$ must fulfil \beq
0=\delta \N_{i\pm}| = [\N_{i\pm} ,K]|+\textstyle{1\over{2(N-2)}}
(1\cdot{\stackrel
\leftarrow K})\N_{i\pm} |.
\eeq{wzcond}
This constrains the various transformation parameters in \hv{Kdef}, and
leads
to the following component relations for $K$:
\ber
\left(1\cdot {\stackrel\leftarrow K}\right)| &=&\half (2-N)\d_{\um}
\s^{\um}+i(N-1)\left(\e^{i+}\h_\+^{\ \ i+} -\e^{i-}\h_=^{\ \
i-}\right)\cr
\N_{i\pm} K| &=& \pm i\e^{i\pm}\N\pp |+\half\e^{i\mp}RM\cr && +\half
\left(i\e^{i-}\h_=^{\ \ i-}-i\e^{i+}\h_\+^{\ \ i+}
\mp\L+\half\d_{\um}\s^{\um}
\right) \d_\pm .
\eer{Kwz}
In particular, for $N=2$, where $(1\cdot{\stackrel\leftarrow K})=0$, we
find
$\h_\+^{\ \ +} = \h_=^{\ \ -}=0$.

Under $(p+1)$-dimensional diffeomorphisms the components transform as
specified by their density weights, and under Lorentz transformations
according to
their Lorentz charge. The local supersymmetry transformations of the
supergravity fields are found from \hv{natra2} using \hv{Kwz}, \hv{nn1},
\hv{nn2} and the component relations. They are;
\ber
\delta e\pp^{\um}&=& \mp i\e^{i\pm}\h\pp^{i\pm} e\pp^{\um}\pm i\e^{i\mp}
(2\h\pp^{i\mp} e\mm^{\um}-\h\mm^{i\mp} e\pp^{\um})\cr \delta \h\pp^{i\pm}
&=& \d\pp\e^{i\pm}
-\e^{i\pm}\left(\half\d_{\um}e\pp^{\um}\pm i\h\pp^{k\mp}\h\mm^{k\mp}\pm
\omega\pp\right)\mp\textstyle{{3i} \over 2}
\e^{j\pm}\h\pp^{j\pm}\h\pp^{i\pm}
\cr
&&+\e^{i\mp}\textstyle{i\over N} R
\pm\e^{j\mp}\left(2i\h\pp^{j\mp}\h\mm^{i\pm}-
\ihalf\h\mm^{j\mp}\h\pp^{i\pm}
\right)\cr \delta\h\pp^{i\mp}&=&\d\pp\e^{i\mp}-\e^{i\mp}\left(\half
\d_{\um}e\pp^{\um}\pm \h\pp^{k\mp}\h\mm^{k\mp}\right)\mp {\textstyle{{3i}
\over 2}}\e^{j\pm}\h\pp^{j\pm}\h\pp^{i\mp}\cr
&&\pm\e^{j\mp}\left(2i\h\pp^{j\mp}\h\mm^{i\mp}-\ihalf\h\mm^{j\mp}\h\pp^{i
\mp}
\right)\cr \delta \omega\pp &=&
-\e^{i\pm}\left(\pm i\h\pp^{i\pm}\omega\pp+\h\pp^{i\mp} R\right)-
\textstyle{{1+N(N-2)}\over{N(N-2)}}\e^{i\mp}\h\pp^{i\pm}R\cr &&-\e^{i\mp}
\left(\mp 2i\h\pp^{i\mp}\omega\mm\pm i\h\mm^{i\mp}\omega\pp\pm
\textstyle{i\over N}\rho_{i\pm}\right),\cr
\delta R &=& -\e^{i+}\rho_{i+} -\e^{i-}
\rho_{i-},
\eer{gtrans}
For completeness, we also present the covariant versions \ber
\delta e_{\ua}^{\ \um} &=& 2i \left(\bar{\e}\g^{\um}\h_{\ua}\right) -i
\left(\bar{\e}\g^{\ub}\h_{\ub}\right)e_{\ua}^{\ \um},\cr \delta
\h_{\ua}^{i\a} &=&\N_{\ua}\e^{i\a} -\textstyle{i\over 2N}R
\left(\bar{\e}^i
\g_{\ua}\right)^\a+2i\left(\bar{\e}\g^{\ub}
\h_{\ua}\right)\h_{\ub}^{i\a}\cr
&&-\ihalf\left(\bar{\e}\g^{\ub}\h_{\ub}\right)\h_{\ua}^{i\a} -i\e^{i\a}
\left(\bar{\h}_{\ua}\g^{\ub}\h_{\ub}\right),\cr
\delta\omega_{\ua} &=&
\textstyle{i\over 2N} \left(\bar{\e}\g_{\ua}\g^5\rho\right)
+i\left(\bar{\e}\g^{\ub}\h_{\ub}\right)\omega_{\ua} + i\left(\bar{\e}
\g_{\ua}\h_{\ub}\right)\omega_{\uc}\e^{\ub\uc}\cr
&& +\textstyle{1\over {2N(N-2)}}\left(\bar{\e}
\g^{\ub}\g_{\ua}\h_{\ub}\right)R 
-\textstyle{{1+N(N-2)}\over{N(N-2)}}\left(\bar{\e}\g^5\h_{\ua}\right)R
,\cr \delta R &=& -\bar{\e}\rho,
\eer{cotf}
where
\beq
\N_{\ua}\e^\a = e_{\ua}^{\ \um}\left(\d_{\um}\e^\a+ \fou \G_{\um\, \un}
^{\un}\e^\a+\omega_{\um} M \e^\a\right),
\eeq{covd}
with $\omega_{\ua}=e_{\ua}^{\ \um}\omega_{\um}$ the full spin-connection,

including torsion,
and $\G_{\um\, \un}^{\up}$ the
$(p+1)$-dimensional connection.
To obtain the supercovariant form of the $\h_{\ua}^{i\a}$ transformation
in
\hv{cotf} we have employed a generalized metricity condition on
$e_{\ua}^{\ \um}$,
\beq
\N_{\um}e_{\ua}^{\ \um}=\d_{\um}e_{\ua}^{\ \um} +\G_{\un\,\um}^{\un}e_
{\ua}^{\ \um}
-\half\G_{\um\,\un}^{\un}e_{\ua}^{\ \um} +\omega_{\um}M e_{\ua}^{\
\um}=0,
\eeq{eeq}
which for $p=1$ is equivalent to the ordinary condition
$\N^{}_{[\um}e_{\un ]}^{\ \ua }=0$.

The matter field transformations are
\ber
\delta X &=& -\e^{i+}\Psi_{i+}-\e^{i-}\Psi_{i-} ,\cr \delta \Psi_{i\pm}
&=&
\mp i\e^{i\pm}\d\pp X -\e^{j+}{\cal F}_{j+i\pm}-\e^{j-}{\cal
F}_{j-i\pm}\cr
&&+\ihalf\left(\e^{j+}\h_{\+}^{\ \ j+}
-\e^{j-}\h_=^{\ \ j-}\right)\P_{i\pm} \cr
&&\mp i\e^{i\pm}\h\pp^{j\pm}\Psi_{j\pm} \mp
i\e^{i\pm}\h\pp^{j\mp}\Psi_{j\mp} .
\eer{Xtfs}
The covariant form of these transformations read \ber
\delta X &=& -\bar{\e}\P,\cr
\delta \P_\a^i &=&i\left(\g^{\um}\e^i\right)_\a \d_{\um}X +i
\left(\g^{\ua}\e^i\right)_\a\left(\bar{\h}_{\ua}\P\right)
\cr &&-\ihalf\left(\bar{\e}\g^{\ua}\h_{\ua}\right)\P_\a^i -
\e^{j\beta}{\cal{F}}_{\beta j,\a i} .
\eer{Xcov}
We will not need $\de {\cal F}$ in general. For $N=1$ it is \ber
\delta {\cal F} &=&
-i\e^-\d_=\Psi_+ -i\e^+\d_\+\Psi_- -\lambda^+\Psi_+ 
-\lambda^-\Psi_- -l^\um\d_\um X \cr
&=& i \bar{\e}\g^{\um}\d_{\um}\P -\bar{\lambda}\P -l^\um\d_\um X ,
 \eer{Ftf}
where $\lambda ^\pm \is \d_+\d_-\e^\pm|$ and $l^\um\is\d_+\d_-\s^\um|$.

Using the transformations in \hv{gtrans} we verify that the
$e_{\ua}^{\ \um}$ constraint \hv{Acons} is supersymmetric.

\section{Actions}

In this section we discuss actions coupling the supergravity to matter
fields. Since the full superspace measure has weight $N/2$, a Lagrangian
has to have weight $1-N/2$. For $N>2$ this becomes negative which cannot
be
achieved using a weight $0$ scalar field and the available operators
(which
have postive weight). In this case one has to resort to integration over
invariant subspaces, using techniques described in , e.g., \cite{proj}.
We
will not treat those cases, and hence the discussion below is restricted
to
$N=1,2$.

A general locally supersymmetric action for $N=1$ is \beq
S=\int d^{p+1}\xi d^2\th{\cal{L}}\left( X,\N X\right). \eeq{smac}
To evaluate the component version of $S$, we need the following relation:
\ber
\int d^{p+1}\xi d^2\th{\cal{L}}&=&\int d^{p+1}\xi
\left(\N_+\N_-+\ihalf\h_\+^{\ \ +}\N_-\right){\cal{L}}|\cr &=&-\int
d^{p+1}\xi
\left(\N_-\N_+-\ihalf\h_=^{\ \ -}\N_+\right){\cal{L}}| \eer{Lcom}

A general $\s$-model action is
\beq
S=\int d^{p+1}\xi d^2\th \, \N_+X^\mu\N_-X^\nu {\cal E}_{\mu\nu}(X),
\eeq{Spinac}
where $\mu , \nu = 1,\ldots,D$ and ${\cal E}_{\mu\nu}\is G_{\mu\nu }+
B_{\mu\nu}$
is the sum of the $D$-dimensional target space metric and antisymmetric
tensor field.
The component version of \hv{Spinac} is\footnote{For simplicity we set
${\cal E}_{\mu\nu}=\eta_{\mu\nu}$. A non trivial ${\cal E}$ will give
rise to target space curvature, connection and torsion terms of 
the usual $\s$-model type.}
\ber
{\bf S}&=&\fou\int
d^{p+1}\xi\left\{\d_\+X\cdot\d_=X+2\h_=^{\ \ +}\d_\+X\cdot\Psi_+ +
2\h_\+^{\ \ -}\d_=X\cdot\Psi_-\right.\cr &&\left.+i\Psi_+\cdot\d_=
\Psi_+-i\Psi_-\cdot\d_\+\Psi_- -2\left(\Psi_+\cdot\Psi_-\right)
\left(\h_=^{\ \ +}\h_\+^{\ \ -}\right)\right.\cr &&\left. +{\cal F}
\cdot{\cal F}
-A_{\um}\d_{\un}(e_{[\+}^{\ \um}e_{=]}^{\ \un})\right\}\cr\cr &=&\fou
\int d^{p+1}\xi
\left\{\eta^{\ua\,\ub}e^{\ \um}_\ua e^{\ \un}_\ub\d_\um X\cdot\d_\un X +
2\bar{\h}_\ua\g^\um\g^\ua\Psi\cdot\d_\um X\right.\cr &&\left.+i\bar{\Psi}
\cdot\g^\um\d_\um\Psi-\half\left({\bar\Psi}\cdot \Psi\right)
\left({\bar\h}_\ua\g^\ub\g^\ua\h_\ub\right)\right.\cr &&\left. +
{\cal F}^{\alpha\beta}\cdot {\cal F}_{\alpha\beta}-\e^{\ua\,\ub}A_\um\d_
\un (e_{\ua}^{\ \um} e_{\ub}^{\ \un})\right\}.
\eer{Spicom}
where we have included the covariant version and taken care of the
constraint $\e^{\ua\,\ub}\d_\un (e_{\ua}^{\ \um} e_{\ub}^{\ \un})=0$ from
\hv{Acons} using a Lagrange multiplier
$A_\um$\footnote{This form of the action is a direct supersymmetrization
of
the strong
coupling limit of the Born-Infeld action, as described in \cite{gulin}.}.
The
last term
is invariant under local supersymmetry provided that $A_\um$ transforms
as a
singlet. To
see the invariance of the action under
diffeomorphisms, note that the only field that is not a density is $X$.

For $N=2$ it is convenient to work with complex objects. We define
\ber
\N_\pm &\is&\N_\pm^1+i\N_\pm^2, \quad \G_\pm \is\G_\pm^1+i\G_\pm^2, \cr
\bN_\pm &\is&\bN_\pm^1-i\bN_\pm^2, \quad \bG_\pm \is\G_\pm^1-i\G_\pm^2,
\eer{complexN}
and
\beq
\h\pp^\a \is \h\pp^{1\a}+i\h\pp^{2\a},\quad
\bh\pp^\a \is \h\pp^{1\a}-i\h\pp^{2\a}.
\eeq{complexchi}
Since for $N=2$ the superdiffeomorphisms are restricted to be
super-volume
preserving, $1\cdot{\stackrel{\leftarrow}{K}} =0$, we must 
put $\h\pp^\pm = \bh\pp^\pm =0$. 
In fact this may be viewed as a superconformal gauge
choice
utilizing the
transformations
\beq
\delta\h\pp^\pm=\eta^\pm,
\eeq{sctf}
where $\eta^\pm$ is
a complex spinor parameter. To stay in this gauge we must require that
the
supersymmetry transformations of $\h\pp^\pm$ be accompanied by
compensating
superconformal transformation \hv{sctf}.

We also introduce ``hatted'' derivatives \beq
\hN_\pm= \N_\pm + 4w\G_\pm , \quad
\hN\pp=\N\pp -2w (1\cdot\stackrel{\leftarrow}{\N}\pp) \eeq{hattedN}
where $w$ is the density weight of the object $\hN$ is acting on. The
constraint algebra then simplifies to that of ordinary $N=2$ supergravity
in $2D$ \cite{howe,jim2}.

Covariantly (anti-)chiral superfields $(\bP)\Phi$ are defined by \beq
\hN _\pm \bP =\hbN_\pm \Phi =0.
\eeq{chiralPhi}
The (hatted covariant) components of the chiral multiplet are defined by
\beq
\begin{array}{lll}
\Phi |= \f , & \hN_\pm\Phi | =\p_\pm , & \hN_+\hN_-\Phi | =F, \\
\bP |= \bfi, & \hbN_\pm \bP |= \bp_\pm , & \hbN_+\hbN_- \bP |= \bF.
\end{array}
\eeq{chiralcomp}
The (anti)chiral measure is $d^2\tb {\cal L}_{chir} =\hbN_+\hbN_-{\cal
L}_{chir}|$. For a general
Lagrangian
${\cal L}$ the
full $N=2$ superspace measure is
\ber
\int d^{p+1}\xi d^4 \th {\cal L} &=&\int d^{p+1}\xi d^2 \th\hbN_+\hbN_-
{\cal L}|_{\tb=0} \cr & =& \int d^{p+1}\xi \left( \hN_+\hN_-
+Y \right) \hbN_+\hbN_- {\cal L}|_{\th =\tb =0}, \eer{chiralmeasure}
where the coefficient $Y$ is determined below.

An action for the chiral multiplet is found by choosing ${\cal
L}=\bP\Phi$,
\ber
{\bf S}&=&{\textstyle{1\over 8}}\int d^{p+1}\xi d^4 \th {\cal L}=\int
d^{p+1}\xi
\left\{ \; 2\d_\+\bfi\d_=\f
\right. \cr
&+&  \textstyle{i\over 4}
\left(\bp_+\d_=\p_+ -\bp_-\d_\+\p_- -\d_=\bp_+\p_+
 +\d_\+\bp_-\p_-\right) \cr 
&+& \bh_\+^{\ \ -}\p_-\d_=\bfi
+\bh_=^{\ \ +}\p_+\d_\+\bfi  
+\h_\+^{\ \ -}\bp_-\d_=\f
+\h_=^{\ \ +}\bp_+\d_\+\f \cr
&-& \left. \half \left( \h_=^{\ \ +}\bh_\+^{\ \ -}\bp_+\p_- 
   +\h_\+^{\ \ -}\bh_=^{\ \ +}\bp_-\p_+\right) 
+\textstyle{1 \over 8}F\bF
-A_{\um}\d_{\un}(e_{[\+}^{\ \um}e_{=]}^{\ \un}) \right\}
\eer{chiralaction}
where we again have incorporated the constraint \hv{Acons} that follows
by
matching the $\d_\ua$ coeffients in \hv{nn}.
Furthermore, in contrast to $N=1$, matching the
$\d_\pm\is\d^1_\pm+i\d^2_\pm$ coefficients we can
solve for $\rho_\pm$ in terms of the other fields,
\ber
{\textstyle{1\over 8}}\r_\pm = \pm\d\mm\h\pp^\mp \pm\half \h\pp^\mp
(\d_\um e\mm^\um ) + \h\pp^\mp\omega\mm,
\eer{rhos}
and similarily for $\br_\pm$.
The $M$ part of \hv{nn} gives
relations for the $\th\tb$-components of $R$,
\ber
&-&\fou\hN_+\hbN_-R| -\fou\hbN_+\hN_-R| -\half R^2 \cr
&=& 2e_{[\+}^{\ \um}\d_\um
\omega^{}_{=]}+4\omega_\+\omega_=
+ \bh_\+^{\ \ -}\h_=^{\ \ +}R-\bh_=^{\ \ +}\h_\+^{\ \ -}R
\eer{hNhNR}

Following \cite{Grisaru} we have determined the coefficient $Y$,
\beq
Y=2\h_\+^{\ \ -}\h_=^{\ \ +} ,
\eeq{Ycoef}
by requiring the terms in the component action \hv{chiralaction}
containing auxiliary fields,
$F$, $\bF$, to be symmetric in barred and unbarred quantities.

The supersymmetry transformations for the chiral- and
antichiral component fields are 
\ber
\delta\f &=& -\half\eb^{\,+}\p_+-\half\eb^{\,-}\p_-, \cr \delta\p_\pm &=&
\mp i\e^\pm\d\pp\f \mp\ihalf\e^\pm\bh\pp^\mp\p_\mp \pm\half\eb^\mp F,
\cr
\delta F &=&
-i\e^-\d_=\p_+-i\e^+\d_\+\p_--\half\bl^+\p_+-\half\bl^-\p_-
-l^\um\d_\um\f,
\eer{Phitf}
and
\ber
\delta\bfi &=& -\half\e^{\,+}\bp_+-\half\e^{\,-}\bp_-, \cr
\delta\bp_\pm &=&
\mp i\eb^\pm\d\pp\bfi \mp\ihalf\eb^\pm\h\pp^\mp\bp_\mp \pm\half\e^\mp
\bF, \cr
\delta \bF &=&
-i\eb^-\d_=\bp_+-i\eb^+\d_\+\bp_--\half\l^+\bp_+-\half\l^-\bp_-
-\bar{l}^\um\d_\um\bfi,
\eer{bPhivtf}
where $\e^\pm=\e_1^\pm+i\e_2^\pm$, $\l^\a =\bd_+\bd_-\e^\a|$
and $l^\um=\d_+\d_-\s^\um|$.

A more general $N=2$ action is
\beq
\int d^{p+1}\xi d^4 \th K(\Phi,\bP).
\eeq{Kaction}
Here the target space geometry is determined by a single potential
function
$K$
leading to a restricted K\"ahler geometry. As is most easily seen from an
analysis of the bosonic content of \hv{Spinac}, \hv{Spicom},
\hv{chiralaction}
and \hv{Kaction}, the non-degenerate
case $p=1$ leads to the usual $2D$ supergravity-matter couplings. The
relation
is via field-redefinitions that reintroduce  the determinant of the
zweibein.

\section{Discussion}

We have presented $[p+1,2]$ supergravities for $N\in\{1,2 \}$. As
mentioned, we could allow for a larger range of $N$, but then the actions
have to be constructed as integrals over invariant subspaces. We may
likewise
extend the treatment to $N=(p,q)$ supergravities. The most direct example
leads to a straightforward generalization of the $(p,0)$ supergravities
of
\cite{Brooks,evans}.

In the previous Section we mentioned that for $p=1$ we recover the
standard
$2D$ supergravities via field redefinitions. We thus have a
novel description of those theories. This description may sometimes be
advantageous, e.g., when discussing the measure in $N=2$.

We find the supergravities presented intrinsically interesting as
examples of
non-standard geometries, but they were developed with one particular
application in mind. The $T\to 0$ limit of the Born-Infeld action
corresponds
to $D$-branes at very large values of the fundamental string coupling. As
shown in \cite{lind1}, the $D$-brane world volume becomes foliated by
string
world
sheets in this limit. Since the discussion in \cite{lind1} is purely
bosonic
and
the fundamental string is supersymmetric, we wanted to confirm this
parton
picture by supersymmetrizing the model. This is presented in
\cite{gulin},
based
on the results reported on here.
\bigskip

{\bf Acknowledgements}
\vskip .2in
\noindent
We are grateful to Martin Ro\v cek for comments and discussions.
The work of UL was supported in part by NFR grant No. F-AA/FU 04038-312
and
by NorFA grant No. 96.55.030-O.
\bigskip

\renewcommand{\thesection}{Appendix}
\setcounter{section}{0}
\renewcommand{\theequation}{\Alph{section}.\arabic{equation}} \setcounter
{equation}{0}

\section{}
In this Appendix we collect some useful relations that were used in the
derivation of the component relations in the text. It also contains our
conventions.

The following (WZ-gauge) relations are needed in evaluating the component
Lagrangian
and in deriving the transformations:
\ber
\N_{i\pm}\N_{j\pm} | &=&\pm i\de_{ij}\left( \h\pp^{l\pm}\d_{l\pm}+
\h\pp^{l\mp}\d_{l\mp}+e\pp^{\um}\d_{\um}+\omega\pp
M \right) \cr
&&\mp\textstyle{i\over{2(2-N)}}\h\pp^{i\pm}\d_{j\pm} +\d_{i\pm}\d_{j\pm},
\cr\cr
\N_{i\pm}\N_{j\mp} | &=&\mp \textstyle{i\over{2(2-N)}} \h\pp^{i\pm}
\d_{j\mp} +\half \de_{ij}R M+\d_{i\pm}\d_{j\mp}
\eer{nn1}
and
\ber
\N\pp\N_{i\pm}|&=&\d\pp\d_{i\pm}+\h\pp^{j+}\N_{j+}\N_{i\pm}|\cr
&&+\h\pp^{j-}\N_{j-}\N_{i\pm}|\pm\half\omega\pp\d_{i\pm},\cr \cr \N\pp
\N_{i\mp}|&=&\d\pp\d_{i\mp}+\h\pp^{j+}\N_{j+}\N_{i\mp}|\cr &&+\h\pp^{j-}
\N_{j-}\N_{i\mp}|\mp\half\omega\pp\d_{i\mp}.
\eer{nn2}

Independent of the gauge, we have the following commutation relations
\ber
\left[\N_{i\pm} ,\N\mm\right] &=&
-2\G_{i\pm}\N\mm \pm\textstyle{i\over N} \left(\N_{j\pm}\G_{j\pm}\right)
\N_{i\pm}\cr &&-\textstyle{i\over{2N}}R\N_{i\mp}\mp
\left(\N_{i\mp}R+\G_{i\mp}R\right)M ,\cr \cr \left[\N_{i\pm},\N\pp\right]
&=&
-\textstyle{{2(N+2)}\over{(N+2)}}\G_{i\pm}\N\pp +\textstyle{i\over N}
\left(\N_{j\pm}\G_{j\pm} \right)\N_{i\pm}
\eer{nN}
and
\ber
\left[\N_\+,\N_= \right] &=& \left\{\textstyle{R\over 2N^2}\G^i_- +
\textstyle{1\over N}(\N^i_+\N^j_-\G^j_-) +\textstyle{1\over N}\G^i_+
(\N^j_-\N^j_-) \right. \cr &+& \left. \textstyle{i\over N}(\N_=\G^i_+) -
\textstyle{1\over 2N^2}(\N^i_-R)\right\}\N^i_+ +\textstyle{2i\over N}
(\N^j_-\G^j_-)\N_\+ \cr
&+& \left\{\textstyle{R\over 2N^2}\G^i_+ -\textstyle{1\over N}
(\N^i_-\N^j_+\G^j_+) -\textstyle{1\over N}\G^i_-(\N^j_+\N^j_+)
\right. \cr &+& \left. \textstyle{i\over N}(\N_\+\G^i_-) -
\textstyle{1\over 2N^2}(\N^i_+R)\right\}\N^i_- +\textstyle{2i\over N}
(\N^j_+\G^j_+)\N_= \cr 
&+& \textstyle{1\over N^2}
\left\{\textstyle{5\over 2}\G^i_{[-}\N^i_{+]}R +6\G^i_-\G^i_+R+R\N^i_{[-}
\G^i_{+]} \right. \cr &-& \left. \half R^2 +\half[\N^i_-,\N^i_+]R\right\}
M .
\eer{nn}

\bigskip

Spinors in a $(p+1)$-dimensional
space-time with non-invertible moving frames $e_{\ua}^{\ \um}$ of rank
$d$,
$\ua=0,\ldots,d-1, \um=0,\ldots,p$ are introduced by prescribing the
Clifford algebra
\beq
\{\g^{\ua},\g^{\ub}\}=2\eta^{\ua\,\ub} \;\;\;\implies\;\;\; \{\g^{\um},
\g^{\un}\}=2g_d^{\um\,\un}\is e_{\ua}^{\ \um}e_{\ub}^{\
\um}\eta^{\ua\,\ub},
\eeq{name}
where $\g^{\um}\is e_{\ua}^{\ \um}\g^{\ua}$ and $\eta^{\ua\,\ub}=(-1,1)$.
For
our case, $d=2$, we use a real representation for the gamma matrices,
$(\g_{\ua})_\a^{\ \beta} =(i\s^2,\s^1)$ and $(\g^5)_\a^{\ \beta}
=(\s^3)$.
The spinor indices are raised and lowered by
$C_{\a\beta}=C^{\a\beta}=i\s^2$,
according to $\h^{\a}=C^{\a\beta}\h_{\beta}$ and $\h_{\a}=\h^{\beta}
C_{\beta\a}$. Since the $2D$ Lorentz group is $SO(1,1)$, which has only
$1$-dimensional representations,
it is convenient to work in a basis of helicity eigenstates. Then a
spinor
index $\a$ takes the values $\{ +,-\}$ (helicity $\pm\half$), and
a (tangent) vector index $\ua$ takes the values $\{ \+,=\}$
(helicity $\pm 1$) and are equivalent to light-cone components:
$(v^{\ua}\g_{\ua})_{\pm,\pm}=\pm v\pp$.
The Lorentz generator $M$ act on spinors and vectors as
\beq
[M,\h_{\pm}] =\pm\half\h_{\pm}, \quad [M,v\pp ]=\pm v\pp.
\eeq{Mact}


\newpage


\begin{thebibliography}{6666}

\newcommand{\np}{{\em Nucl.\ Phys.\ }}
\newcommand{\pr}{{\em Phys.\ Rev.\ }}
\newcommand{\cmp}{{\em Commun.\ Math.\ Phys.\ }} \newcommand{\pl}
{{\em Phys.\ Lett.\ }}
\newcommand{\prl}{{\em Phys.\ Rev.\ Lett.\ }} \newcommand{\cqg}
{{\em Class.\ Quantum\ Grav.\ }} \bibitem{witt1}
E. Witten, \np {\bf B311} (1988) 46

\bibitem{3dgr}
J. H. Horne and E. Witten, \prl {\bf 62} (1989 ) 501; U. Lindstr\"om and
M. Ro\v cek \prl {\bf 62} (1989) 2905.

\bibitem{kl}
A.\ Karlhede and U.\ Lindstr\"om, \cqg {\bf 3} (1986) L73.

\bibitem{lst2}
U.\ Lindstr\"om,\ B.\ Sundborg and G.\ Theodoridis \pl {\bf 253B} (1991)
319.

\bibitem{lst1}
U.\ Lindstr\"om,\ B.\ Sundborg and G.\ Theodoridis \pl {\bf 258B} (1991)
331.

\bibitem{lind1}
U.\ Lindstr\"om and R.\ von Unge, \pl {\bf B403} (1997) 233.

\bibitem{lr}
U. Lindstr\"om and M. Ro\v cek, \pl {\bf 271B} (1991) 79.

\bibitem{jim}
S. J. Gates Jr. and H. Nishino, \cqg {\bf 3} (1986) 391.

\bibitem{rvnz}
M. Ro\v cek, P. van Nieuwenhuizen and S. C. Zhang, {\em Annals of Phys.}
{\bf 172} (1985) 348.

\bibitem{howe}
P. S. Howe and G. Papadopoulos, \cqg {\bf 4} (1987) 11.

\bibitem{jim2}
S. J. Gates Jr., L. Liu and N. Oerter, \pl {\bf 218B} (1989) 33.

\bibitem{proj}
U.\ Lindstr\"om and M.\ Ro\v cek, \cmp {\bf 128} (1990) 191.


\bibitem{gulin}
H.\ Gustafsson and U.\ Lindstr\"om, ``A Picture of $D$-branes at strong
coupling II. Spinning Partons.'' University of Stockholm preprint
USITP-98-13, (1998),
hep-th/9807064.


\bibitem{Grisaru}
M. T. Grisaru and M. E. Wehlau, \np {\bf B457} (1995) 219.

\bibitem{Brooks}
R. Brooks, F. Muhammad and S. J. Gates Jr., \np {\bf B268} (1986) 599.

\bibitem{evans}
M. Evans and B. A. Ovrut, \pl {\bf 186B} (1987) 134.


\end{thebibliography}
\end{document}